\documentclass[10pt]{article}
\usepackage{graphicx}
\usepackage{amsmath}
\usepackage{amssymb}
\usepackage{caption2}
\begin{document}
\bibliographystyle{prsty}
\begin{center}
{\large {\bf \sc{  Tetraquark candidates in the  LHCb's  di-$J/\psi$ mass spectrum }}} \\[2mm]
Zhi-Gang  Wang \footnote{E-mail: zgwang@aliyun.com.  }   \\
 Department of Physics, North China Electric Power University, Baoding 071003, P. R. China
\end{center}

\begin{abstract}
In this article, we study the first radial excited states of the  scalar, axialvector, vector and tensor  diquark-antidiquark type $cc\bar{c}\bar{c}$  tetraquark states with the QCD sum rules and obtain the masses and pole residues, then we use the Regge trajectory to obtain the masses of the second radial excited states. The predicted masses  support assigning the broad structure from 6.2 to 6.8 GeV in the di-$J/\psi$ mass spectrum to be the first radial excited state of the scalar, axialvector, vector or tensor $cc\bar{c}\bar{c}$ tetraquark state, and assigning the narrow structure at about 6.9 GeV in the di-$J/\psi$ mass spectrum to be the  second radial excited state of the scalar or axialvector $cc\bar{c}\bar{c}$ tetraquark state.
\end{abstract}



\section{Introduction}
The  charmonium-like and bottomonium-like  states  are good subjects   to study the  exotic states and understand the strong interactions, if they are  genuine tetraquark states, there are two heavy valence quarks and two light valence  quarks, therefore the dynamics is complex compared to the tetraquark  configurations consist of four heavy valence quarks, the attractive interactions between the two heavy quarks (or antiquarks) should dominate at the short distance and favor
forming the genuine diquark-antidiquark type tetraquark states rather
than the loosely bound tetraquark molecular states, because the light mesons  cannot be exchanged between the two heavy quarkonia to provide attractions  at the leading order.
In recent years, the full-heavy tetraquark states have attracted much attentions and have been studied extensively \cite{Lloyd-2004,Barnea-2006,Berezhnoy-2012,Heupel-2012,Bai-2016,Richard-2017,WZG-QQQQ,Rosner-2017,Chen-2017,GuoFK-bbbb,
Polosa-2018,Wu-2018,Hughes-2018,Navarra-2019,Zhong-2019,Chen-2019,Bedolla-2019,Ping-2020,Ohlsson-2020,Zhong-2020}.

Recently, the LHCb collaboration reported their preliminary results on the observations of the $cc\bar{c}\bar{c}$ tetraquark  candidates  in the di-$J/\psi$ invariant mass spectrum  at $p_T> 5.2\,\rm{GeV}$   \cite{LHCb-cccc}. They observed  a broad structure above the threshold ranging from 6.2 to 6.8 GeV and a narrow structure at about 6.9 GeV  with the significance of more than $5\sigma$, furthermore, they also observed  some vague structures around 7.2 GeV.
 The masses of the full-heavy   tetraquark states from the  phenomenological quark models lie either  above or below the di-$J/\psi$ or di-$\Upsilon$ threshold, and vary at a large range \cite{Lloyd-2004,Barnea-2006,Berezhnoy-2012,Heupel-2012,Bai-2016,Richard-2017,WZG-QQQQ,Rosner-2017,Chen-2017,GuoFK-bbbb,
Polosa-2018,Wu-2018,Hughes-2018,Navarra-2019,Zhong-2019,Chen-2019,Bedolla-2019,Ping-2020,Ohlsson-2020,Zhong-2020}.
 It is the first time that  clear structures in the di-$J/\psi$ mass spectrum  are observed experimentally, which may be evidences for the genuine  $cc\bar{c}\bar{c}$ tetraquark states.
  The observation of evidences for the $cc\bar{c}\bar{c}$ tetraquark states
   provides  important experimental constraints on the theoretical models and sheds  light on the nature of  the exotic states, and plays
an important role in establishing  the tetraquark states.

In Ref.\cite{WZG-QQQQ}, we study the mass spectrum of the ground states of the scalar, axialvector, vector and tensor full-heavy diquark-antidiquark type tetraquark states with the QCD sum rules, the predicted tetraquark masses lie blow the di-$J/\psi$ or di-$\Upsilon$ threshold. In the present work, we extend our previous work to study the  mass spectrum of the first radial excited  states of the scalar, axialvector, vector and tensor diquark-antidiquark type $cc\bar{c}\bar{c}$ tetraquark states with the QCD sum rules, then take the masses of the ground states and the first radial excited states as the input parameters, resort to the Regge trajectory to obtain the masses of the second radial excited states, and make possible assignments of the LHCb's new structures.

The article is arranged as follows:  we derive the QCD sum rules for the masses and pole residues of  the first radial excited states of the  $cc\bar{c}\bar{c}$ tetraquark states in section 2; in section 3, we present the numerical results and use the Regge trajectory to obtain the masses of the second  radial excited states; section 4 is reserved for our conclusion.

\section{QCD sum rules for  the first radial excited  $cc\bar{c}\bar{c}$ tetraquark states  }
Let us  write down  the two-point correlation functions   $\Pi (p)$ and $\Pi_{\mu\nu\alpha\beta}(p)$ in the QCD sum rules firstly,
\begin{eqnarray}
\Pi(p)&=&i\int d^4x e^{ip \cdot x} \langle0|T\left\{J(x)J^{\dagger}(0)\right\}|0\rangle \, ,\nonumber\\
\Pi_{\mu\nu\alpha\beta}(p)&=&i\int d^4x e^{ip \cdot x} \langle0|T\left\{J_{\mu\nu}(x)J_{\alpha\beta}^{\dagger}(0)\right\}|0\rangle \, ,
\end{eqnarray}
where $J_{\mu\nu}(x)=J^1_{\mu\nu}(x)$, $J^2_{\mu\nu}(x)$,
\begin{eqnarray}
J(x)&=&\varepsilon^{ijk}\varepsilon^{imn}c^{Tj}(x)C\gamma_\mu c^k(x) \bar{c}^m(x)\gamma^\mu C \bar{c}^{Tn}(x) \, , \nonumber\\
J^1_{\mu\nu}(x)&=&\varepsilon^{ijk}\varepsilon^{imn}\Big\{c^{Tj}(x)C\gamma_\mu c^k(x) \bar{c}^m(x) \gamma_\nu C \bar{c}^{Tn}(x)-c^{Tj}(x)C\gamma_\nu  c^k(x) \bar{c}^m(x)\gamma_\mu C \bar{c}^{Tn}(x) \Big\} \, , \nonumber \\
J^2_{\mu\nu}(x)&=&\frac{\varepsilon^{ijk}\varepsilon^{imn}}{\sqrt{2}}\Big\{c^{Tj}(x)C\gamma_\mu c^k(x) \bar{c}^m(x) \gamma_\nu C \bar{c}^{Tn}(x)+c^{Tj}(x)C\gamma_\nu  c^k(x) \bar{c}^m(x)\gamma_\mu C \bar{c}^{Tn}(x) \Big\} \, , \nonumber \\
\end{eqnarray}
 the $i$, $j$, $k$, $m$, $n$ are color indexes, the $C$ is the charge conjunction matrix. We choose  the   currents $J(x)$, $J^1_{\mu\nu}(x)$ and $J^2_{\mu\nu}(x)$ to interpolate the $J^{PC}=0^{++}$, $1^{+-}$, $1^{--}$ and $2^{++}$ diquark-antidiquark type $cc\bar{c}\bar{c}$ tetraquark states, respectively, as the current $J^1_{\mu\nu}(x)$, where the Lorentz indexes $\mu$  and $\nu$ are antisymmetric,  has both the spin-parity $J^P=1^+$ and $1^-$ components.

At the hadron  side, we  insert  a complete set of intermediate hadronic states with
the same quantum numbers as the current operators $J(x)$, $J^1_{\mu\nu}(x)$ and $J^2_{\mu\nu}(x)$ into the
correlation functions $\Pi(p)$ and $\Pi_{\mu\nu\alpha\beta}(p)$ to obtain the hadronic representation
\cite{SVZ79,Reinders85}. After isolating the ground state
contributions of the scalar, axialvector, vector  and tensor $cc\bar{c}\bar{c}$  tetraquark states, we obtain the results,  
\begin{eqnarray}
\Pi (p) &=&\frac{\lambda_X^2}{M^2_X-p^2} +\cdots \, \, , \nonumber\\
&=&\Pi_S(p^2)\, ,
\end{eqnarray}
\begin{eqnarray}
\Pi^1_{\mu\nu\alpha\beta}(p)&=&\frac{\lambda_{ Y^+}^2}{M_{Y^+}^2\left(M_{Y^+}^2-p^2\right)}\left(p^2g_{\mu\alpha}g_{\nu\beta} -p^2g_{\mu\beta}g_{\nu\alpha} -g_{\mu\alpha}p_{\nu}p_{\beta}-g_{\nu\beta}p_{\mu}p_{\alpha}+g_{\mu\beta}p_{\nu}p_{\alpha}+g_{\nu\alpha}p_{\mu}p_{\beta}\right) \nonumber\\
&&+\frac{\lambda_{ Y^-}^2}{M_{Y^-}^2\left(M_{Y^-}^2-p^2\right)}\left( -g_{\mu\alpha}p_{\nu}p_{\beta}-g_{\nu\beta}p_{\mu}p_{\alpha}+g_{\mu\beta}p_{\nu}p_{\alpha}+g_{\nu\alpha}p_{\mu}p_{\beta}\right) +\cdots \, \, ,\nonumber\\
&=&\Pi_{A}(p^2)\left(p^2g_{\mu\alpha}g_{\nu\beta} -p^2g_{\mu\beta}g_{\nu\alpha} -g_{\mu\alpha}p_{\nu}p_{\beta}-g_{\nu\beta}p_{\mu}p_{\alpha}+g_{\mu\beta}p_{\nu}p_{\alpha}+g_{\nu\alpha}p_{\mu}p_{\beta}\right) \nonumber\\
&&+\Pi_{V}(p^2)\left( -g_{\mu\alpha}p_{\nu}p_{\beta}-g_{\nu\beta}p_{\mu}p_{\alpha}+g_{\mu\beta}p_{\nu}p_{\alpha}+g_{\nu\alpha}p_{\mu}p_{\beta}\right) \, .
\end{eqnarray}
\begin{eqnarray}
\Pi^2_{\mu\nu\alpha\beta} (p) &=&\frac{\lambda_X^2}{M_X^2-p^2}\left( \frac{\widetilde{g}_{\mu\alpha}\widetilde{g}_{\nu\beta}+\widetilde{g}_{\mu\beta}\widetilde{g}_{\nu\alpha}}{2}-\frac{\widetilde{g}_{\mu\nu}\widetilde{g}_{\alpha\beta}}{3}\right) +\cdots \, \, ,  \nonumber\\
&=&\Pi_{T}(p^2)\left( \frac{\widetilde{g}_{\mu\alpha}\widetilde{g}_{\nu\beta}+\widetilde{g}_{\mu\beta}\widetilde{g}_{\nu\alpha}}{2}-\frac{\widetilde{g}_{\mu\nu}\widetilde{g}_{\alpha\beta}}{3}\right) +\cdots \, \, ,
\end{eqnarray}
where $\widetilde{g}_{\mu\nu}=g_{\mu\nu}-\frac{p_{\mu}p_{\nu}}{p^2}$, the pole residues  $\lambda_{X}$ and $\lambda_{Y}$ are defined by
\begin{eqnarray}
 \langle 0|J (0)|X (p)\rangle &=& \lambda_{X}     \, , \nonumber\\
  \langle 0|J^1_{\mu\nu}(0)|Y^+(p)\rangle &=& \frac{\lambda_{Y^+}}{M_{Y^+}} \, \varepsilon_{\mu\nu\alpha\beta} \, \varepsilon^{\alpha}p^{\beta}\, , \nonumber\\
 \langle 0|J^1_{\mu\nu}(0)|Y^-(p)\rangle &=& \frac{\lambda_{Y^-}}{M_{Y^-}} \left(\varepsilon_{\mu}p_{\nu}-\varepsilon_{\nu}p_{\mu} \right)\, ,\nonumber\\
  \langle 0|J^2_{\mu\nu}(0)|X (p)\rangle &=& \lambda_{X} \, \varepsilon_{\mu\nu}   \, ,
\end{eqnarray}
the $\varepsilon_{\mu}$ and $\varepsilon_{\mu\nu} $ are the  polarization vectors of the axialvector, vector and tensor tetraquark states, respectively.

If we take into account (or isolate) the first radial excited states, we obtain
\begin{eqnarray}
\Pi_{S/T}(p^2)&=&\frac{\lambda_X^2}{M^2_X-p^2} +\frac{\lambda_{X^\prime}^2}{M^2_{X^\prime}-p^2}+\cdots \, , \nonumber\\
\Pi_{A/V}(p^2)&=&\frac{\lambda_{ Y^\pm}^2}{M_{Y^\pm}^2\left(M_{Y^\pm}^2-p^2\right)}+\frac{\lambda_{ Y^{\prime\pm}}^2}{M_{Y^{\prime\pm}}^2\left(M_{Y^{\prime\pm}}^2-p^2\right)}+\cdots\, .
\end{eqnarray}
We project out the axialvector and vector components $\Pi_{A}(p^2)$ and $\Pi_{V}(p^2)$ by introducing the operators $P_{A}^{\mu\nu\alpha\beta}$ and $P_{V}^{\mu\nu\alpha\beta}$, respectively, 
\begin{eqnarray}
\widetilde{\Pi}_{A}(p^2)&=&p^2\Pi_{A}(p^2)=P_{A}^{\mu\nu\alpha\beta}\Pi_{\mu\nu\alpha\beta}(p) \, , \nonumber\\
\widetilde{\Pi}_{V}(p^2)&=&p^2\Pi_{V}(p^2)=P_{V}^{\mu\nu\alpha\beta}\Pi_{\mu\nu\alpha\beta}(p) \, ,
\end{eqnarray}
where
\begin{eqnarray}
P_{A}^{\mu\nu\alpha\beta}&=&\frac{1}{6}\left( g^{\mu\alpha}-\frac{p^\mu p^\alpha}{p^2}\right)\left( g^{\nu\beta}-\frac{p^\nu p^\beta}{p^2}\right)\, , \nonumber\\
P_{V}^{\mu\nu\alpha\beta}&=&\frac{1}{6}\left( g^{\mu\alpha}-\frac{p^\mu p^\alpha}{p^2}\right)\left( g^{\nu\beta}-\frac{p^\nu p^\beta}{p^2}\right)-\frac{1}{6}g^{\mu\alpha}g^{\nu\beta}\, .
\end{eqnarray}

It is straightforward but tedious to carry out the operator product expansion at the deep Euclidean  space $P^2=-p^2 \to \infty$ or $\gg \Lambda_{QCD}^2$, then we obtain the QCD spectral densities through dispersion relation \cite{WZG-QQQQ},
\begin{eqnarray}
\Pi_{S/T}(p^2)&=& \int_{16m_c^2}^{\infty}ds \frac{\rho_{S/T}(s)}{s-p^2}\, ,\nonumber\\
\widetilde{\Pi}_{A/V}(p^2)&=& \int_{16m_c^2}^{\infty}ds \frac{\rho_{A/V}(s)}{s-p^2}\, ,
\end{eqnarray}
where
\begin{eqnarray}
\rho_{S/T}(s)&=&\frac{{\rm Im}\Pi_{S/T}(s)}{\pi}\, , \nonumber\\
\rho_{A/V}(s)&=&\frac{{\rm Im}\widetilde{\Pi}_{A/V}(s)}{\pi}\, .
\end{eqnarray}

 We  take the quark-hadron duality below the continuum thresholds  $s_0$ and $s_0^\prime$, respectively,  and perform Borel transform  with respect to
the variable $P^2=-p^2$ to obtain  the QCD sum rules:
\begin{eqnarray}\label{QCDST-1S}
\lambda^2_{X/Y}\, \exp\left(-\frac{M^2_{X/Y}}{T^2}\right)&=& \int_{16m_c^2}^{s_0} ds \int_{z_i}^{z_f}dz \int_{t_i}^{t_f}dt \int_{r_i}^{r_f}dr\, \rho(s,z,t,r)  \exp\left(-\frac{s}{T^2}\right) \, ,
\end{eqnarray}
\begin{eqnarray}\label{QCDST-2S}
\lambda^2_{X/Y}\, \exp\left(-\frac{M^2_{X/Y}}{T^2}\right)+\lambda^2_{X^\prime/Y^\prime}\, \exp\left(-\frac{M^2_{X^\prime/Y^\prime}}{T^2}\right)&=& \int_{16m_c^2}^{s^\prime_0} ds \int_{z_i}^{z_f}dz \int_{t_i}^{t_f}dt \int_{r_i}^{r_f}dr\, \rho(s,z,t,r)  \nonumber\\
&& \exp\left(-\frac{s}{T^2}\right) \, ,
\end{eqnarray}
where the QCD spectral densities  $\rho(s,z,t,r) =\rho_S(s,z,t,r) $, $\rho_A(s,z,t,r) $, $\rho_V(s,z,t,r)$ and $\rho_T(s,z,t,r) $,
\begin{eqnarray}
\rho_S(s,z,t,r)&=& \frac{3m_c^4}{8\pi^6}\left( s-\overline{m}_c^2\right)^2+\frac{t z m_c^2}{8\pi^6}\left( s-\overline{m}_c^2\right)^2\left( 5s-2\overline{m}_c^2\right) \nonumber\\
&&+\frac{rtz(1-r-t-z)}{1-t-z} \frac{1}{32\pi^6}\left( s-\overline{m}_c^2\right)^3\left( 3s-\overline{m}_c^2\right) \nonumber\\
&&+\frac{rtz(1-r-t-z)}{1-z} \frac{1}{32\pi^6}\left( s-\overline{m}_c^2\right)^3\left( 3s-\overline{m}_c^2\right)\left[5-\frac{t}{1-t-z} \right] \nonumber\\
&&-\frac{rtz^2(1-r-t-z)}{1-z} \frac{3}{16\pi^6}\left( s-\overline{m}_c^2\right)^4 \nonumber\\
&&+rtz(1-r-t-z) \frac{3s}{8\pi^6}\left( s-\overline{m}_c^2\right)^2\left[ 2s-\overline{m}_c^2-\frac{z}{1-z}\left( s-\overline{m}_c^2\right)\right] \nonumber\\
&&+m_c^2\langle \frac{\alpha_sGG}{\pi}\rangle \left\{-\frac{1}{r^3} \frac{m_c^4}{6\pi^4}\delta\left( s-\overline{m}_c^2\right) -\frac{1-r-t-z}{r^2} \frac{m_c^2}{12\pi^4}\left[2+s\,\delta\left( s-\overline{m}_c^2\right)\right]\right. \nonumber\\
&&-\frac{tz}{r^3} \frac{m_c^2}{12\pi^4}\left[2+s\,\delta\left( s-\overline{m}_c^2\right)\right]  -\frac{tz(1-r-t-z)}{r^2(1-t-z)} \frac{1}{12\pi^6}\left( 3s-2\overline{m}_c^2\right) \nonumber\\
&&-\frac{tz(1-r-t-z)}{r^2(1-z)} \frac{1}{12\pi^4}\left( 3s-2\overline{m}_c^2\right) \left[5-\frac{t}{1-t-z} \right]\nonumber\\
&&+\frac{tz^2(1-r-t-z)}{r^2(1-z)} \frac{1}{\pi^4}\left( s-\overline{m}_c^2\right)  \nonumber\\
&&-\frac{tz(1-r-t-z)}{r^2} \frac{1}{2\pi^4}\left[s+\frac{s^2}{3}\delta\left( s-\overline{m}_c^2\right)-\frac{z}{1-z}s\right]  \nonumber\\
&&\left.+\frac{1}{r^2} \frac{m_c^2}{2\pi^4}   +\frac{tz}{r^2} \frac{1}{4\pi^4}\left( 3s-2\overline{m}_c^2\right)-  \frac{1}{16\pi^4}\left( 3s-2\overline{m}_c^2\right)  \right\} \nonumber
\end{eqnarray}
\begin{eqnarray}
&&+\langle \frac{\alpha_sGG}{\pi}\rangle \left\{\frac{1}{rz} \frac{m_c^4}{6\pi^4} +\frac{t}{r} \frac{m_c^2}{6\pi^4}\left( 3s-2\overline{m}_c^2\right)\right. \nonumber\\
&&+\frac{t(1-r-t-z)}{(1-t-z)} \frac{1}{12\pi^4}\left( s-\overline{m}_c^2\right) \left( 2s-\overline{m}_c^2\right)  \nonumber\\
&&+\frac{t(1-r-t-z)}{(1-z)} \frac{1}{12\pi^4}\left( s-\overline{m}_c^2\right) \left( 2s-\overline{m}_c^2\right) \left[2-\frac{t}{1-t-z} \right] \nonumber\\
&&-\frac{tz(1-r-t-z)}{(1-z)} \frac{1}{4\pi^4}\left( s-\overline{m}_c^2\right)^2\nonumber\\
&&\left.+t(1-r-t-z) \frac{1}{12\pi^4}s\left[ 4s-3\overline{m}_c^2-\frac{z}{1-z}3\left( s-\overline{m}_c^2\right)\right]\right\}\, ,
\end{eqnarray}

\begin{eqnarray}
\rho_T(s,z,t,r)&=& \frac{3m_c^4}{16\pi^6}\left( s-\overline{m}_c^2\right)^2+\frac{t z m_c^2}{8\pi^6}\left( s-\overline{m}_c^2\right)^2\left( 4s-\overline{m}_c^2\right) \nonumber\\
&&+\frac{rtz(1-r-t-z)}{1-t-z} \frac{1}{320\pi^6}\left( s-\overline{m}_c^2\right)^3\left( 17s-5\overline{m}_c^2\right) \nonumber\\
&&+\frac{rtz(1-r-t-z)}{1-z} \frac{1}{320\pi^6}\left( s-\overline{m}_c^2\right)^3\left[\left( 21s-5\overline{m}_c^2\right)-\frac{t}{1-t-z}\left( 17s-5\overline{m}_c^2\right) \right] \nonumber\\
&&-\frac{rtz^2(1-r-t-z)}{1-z} \frac{1}{32\pi^6}\left( s-\overline{m}_c^2\right)^4 \nonumber\\
&&+rtz(1-r-t-z) \frac{s}{80\pi^6}\left( s-\overline{m}_c^2\right)^2\left[ 28s-13\overline{m}_c^2-\frac{z}{1-z}7\left( s-\overline{m}_c^2\right)\right] \nonumber\\
&&+m_c^2\langle \frac{\alpha_sGG}{\pi}\rangle \left\{-\frac{1}{r^3} \frac{m_c^4}{12\pi^4}\delta\left( s-\overline{m}_c^2\right) -\frac{1-r-t-z}{r^2} \frac{m_c^2}{12\pi^4}\left[1+s\,\delta\left( s-\overline{m}_c^2\right)\right]\right. \nonumber\\
&&-\frac{tz}{r^3} \frac{m_c^2}{12\pi^4}\left[1+s\,\delta\left( s-\overline{m}_c^2\right)\right]  -\frac{tz(1-r-t-z)}{r^2(1-t-z)} \frac{1}{12\pi^6}\left( 2s-\overline{m}_c^2\right) \nonumber\\
&&-\frac{tz(1-r-t-z)}{r^2(1-z)} \frac{1}{12\pi^4}\left( 2s-\overline{m}_c^2\right) \left[1-\frac{t}{1-t-z} \right]\nonumber\\
&&+\frac{tz^2(1-r-t-z)}{r^2(1-z)} \frac{1}{6\pi^4}\left( s-\overline{m}_c^2\right)  \nonumber\\
&&-\frac{tz(1-r-t-z)}{r^2} \frac{1}{6\pi^4}\left[s+\frac{s^2}{2}\delta\left( s-\overline{m}_c^2\right)-\frac{z}{1-z}s\right]  \nonumber\\
&&\left.+\frac{1}{r^2} \frac{m_c^2}{4\pi^4}   +\frac{tz}{r^2} \frac{1}{4\pi^4}\left( 2s-\overline{m}_c^2\right)  \right\} \nonumber
\end{eqnarray}
\begin{eqnarray}
&&+\langle \frac{\alpha_sGG}{\pi}\rangle \left\{-  \frac{m_c^2}{48\pi^4}\left( 4s-3\overline{m}_c^2\right)\right. \nonumber\\
&&-\frac{r(1-r-t-z)}{1-t-z}\frac{1}{32\pi^4} \left( s-\overline{m}_c^2\right)\left( 3s-\overline{m}_c^2\right)\nonumber\\
&&-\frac{r(1-r-t-z)}{1-z}\frac{1}{480\pi^4} \left( s-\overline{m}_c^2\right)\left[\left( 17s-5\overline{m}_c^2\right)-\frac{t}{1-t-z}15\left( 3s-\overline{m}_c^2\right)\right]\nonumber\\
&&+\frac{rz(1-r-t-z)}{1-z}\frac{1}{24\pi^4} \left( s-\overline{m}_c^2\right)^2 \nonumber\\
&&-r(1-r-t-z)\frac{1}{240\pi^4} s\left[\left( 14s-9\overline{m}_c^2\right)-\frac{z}{1-z}21\left( s-\overline{m}_c^2\right)\right]\nonumber\\
&&-\frac{1}{rz} \frac{m_c^4}{36\pi^4} -\frac{t}{r} \frac{m_c^2}{18\pi^4}\left( 2s-\overline{m}_c^2\right)\nonumber\\
&&-\frac{t(1-r-t-z)}{(1-t-z)} \frac{1}{72\pi^4}\left( s-\overline{m}_c^2\right) \left( 4s-\overline{m}_c^2\right)  \nonumber\\
&&-\frac{t(1-r-t-z)}{(1-z)} \frac{1}{72\pi^4}\left( s-\overline{m}_c^2\right)  \left[2\left( 2s-\overline{m}_c^2\right)-\frac{t}{1-t-z}\left( 4s-\overline{m}_c^2\right) \right] \nonumber\\
&&+\frac{tz(1-r-t-z)}{(1-z)} \frac{1}{24\pi^4}\left( s-\overline{m}_c^2\right)^2\nonumber\\
&&\left.-t(1-r-t-z) \frac{1}{72\pi^4}s\left[ 7s-5\overline{m}_c^2-\frac{z}{1-z}5\left( s-\overline{m}_c^2\right)\right]\right\}\, ,
\end{eqnarray}

\begin{eqnarray}
\rho_A(s,z,t,r)&=& \frac{3m_c^4}{16\pi^6}\left( s-\overline{m}_c^2\right)^2+\frac{t z m_c^2}{8\pi^6}\left( s-\overline{m}_c^2\right)^2\left( 4s-\overline{m}_c^2\right) \nonumber\\
&&+rtz(1-r-t-z) \frac{s}{16\pi^6}\left( s-\overline{m}_c^2\right)^2\left( 7s-4\overline{m}_c^2\right) \nonumber\\
&&+m_c^2\langle \frac{\alpha_sGG}{\pi}\rangle \left\{-\frac{1}{r^3} \frac{m_c^4}{12\pi^4}\delta\left( s-\overline{m}_c^2\right) -\frac{1-r-t-z}{r^2} \frac{m_c^2}{12\pi^4}\left[1+s\,\delta\left( s-\overline{m}_c^2\right)\right]\right. \nonumber\\
&&-\frac{tz}{r^3} \frac{m_c^2}{12\pi^4}\left[1+s\,\delta\left( s-\overline{m}_c^2\right)\right] -\frac{tz(1-r-t-z)}{r^2} \frac{1}{12\pi^4}\left[4s+s^2\delta\left( s-\overline{m}_c^2\right)\right]  \nonumber\\
&&\left.+\frac{1}{r^2} \frac{m_c^2}{4\pi^4}   +\frac{tz}{r^2} \frac{1}{4\pi^4}\left( 2s-\overline{m}_c^2\right) \right\} \nonumber\\
&&+\langle \frac{\alpha_sGG}{\pi}\rangle \left\{-\frac{m_c^2}{48\pi^4}\left( 4s-3\overline{m}_c^2\right)- \frac{r(1-r-t-z)}{16\pi^4}\left( s-\overline{m}_c^2\right)^2\right. \nonumber\\
&&- \frac{r(1-r-t-z)}{48\pi^4}s\left( 7s-6\overline{m}_c^2\right)+\frac{1}{rz} \frac{m_c^4}{48\pi^4} +\frac{t}{r} \frac{m_c^2}{24\pi^4}\left( 2s-\overline{m}_c^2\right)\nonumber\\
&&\left.+ \frac{t(1-r-t-z)}{32\pi^4}\left( s-\overline{m}_c^2\right)^2+ \frac{t(1-r-t-z)}{48\pi^4}s\left( 6s-5\overline{m}_c^2\right)\right\}\, ,
\end{eqnarray}

\begin{eqnarray}
\rho_V(s,z,t,r)&=& -\frac{3m_c^4}{16\pi^6}\left( s-\overline{m}_c^2\right)^2-\frac{t z m_c^2}{8\pi^6}\left( s-\overline{m}_c^2\right)^3 \nonumber\\
&&+rtz(1-r-t-z) \frac{s}{16\pi^6}\left( s-\overline{m}_c^2\right)^2\left( 7s-4\overline{m}_c^2\right) \nonumber\\
&&+m_c^2\langle \frac{\alpha_sGG}{\pi}\rangle \left\{\frac{1}{r^3} \frac{m_c^4}{12\pi^4}\delta\left( s-\overline{m}_c^2\right)+\frac{1-r-t-z}{r^2} \frac{m_c^2}{12\pi^4}\right. \nonumber\\
&&+\frac{tz}{r^3} \frac{m_c^2}{12\pi^4} -\frac{tz(1-r-t-z)}{r^2} \frac{1}{12\pi^4}\left[4s+s^2\delta\left( s-\overline{m}_c^2\right)\right]  \nonumber\\
&&\left.-\frac{1}{r^2} \frac{m_c^2}{4\pi^4}   -\frac{tz}{r^2} \frac{1}{4\pi^4}\left( s-\overline{m}_c^2\right) \right\} \nonumber\\
&&+\langle \frac{\alpha_sGG}{\pi}\rangle \left\{\frac{m_c^2}{48\pi^4}\left( 5s-3\overline{m}_c^2\right)+ \frac{r(1-r-t-z)}{16\pi^4}\left( s-\overline{m}_c^2\right)^2\right. \nonumber\\
&&+ \frac{r(1-r-t-z)}{48\pi^4}s\left( 7s-6\overline{m}_c^2\right)-\frac{1}{rz} \frac{m_c^4}{48\pi^4} -\frac{t}{r} \frac{m_c^2}{24\pi^4}\left( s-\overline{m}_c^2\right)\nonumber\\
&&\left.- \frac{t(1-r-t-z)}{32\pi^4}\left( s-\overline{m}_c^2\right)^2- \frac{t(1-r-t-z)}{48\pi^4}s\left( s-\overline{m}_c^2\right)\right\}\, ,
\end{eqnarray}
and
\begin{eqnarray}
\overline{m}_c^2&=&\frac{m_c^2}{r}+\frac{m_c^2}{t}+\frac{m_c^2}{z}+\frac{m_c^2}{1-r-t-z}\, ,\nonumber 
\end{eqnarray}
\begin{eqnarray}
r_{f/i}&=&\frac{1}{2}\left\{1-z-t \pm \sqrt{(1-z-t)^2-4\frac{1-z-t}{\hat{s}-\frac{1}{z}-\frac{1}{t}}}\right\} \, ,\nonumber\\
t_{f/i}&=&\frac{1}{2\left( \hat{s}-\frac{1}{z}\right)}\left\{ (1-z)\left( \hat{s}-\frac{1}{z}\right)-3 \pm \sqrt{ \left[  (1-z)\left( \hat{s}-\frac{1}{z}\right)-3\right]^2-4 (1-z)\left( \hat{s}-\frac{1}{z}\right)  }\right\}\, ,\nonumber\\
z_{f/i}&=&\frac{1}{2\hat{s}}\left\{ \hat{s}-8 \pm \sqrt{\left(\hat{s}-8\right)^2-4\hat{s}  }\right\}\, ,
\end{eqnarray}
and $\hat{s}=\frac{s}{m_c^2}$. We   introduce the notations $\tau=\frac{1}{T^2}$, $D^n=\left( -\frac{d}{d\tau}\right)^n$, and  use the subscripts $1$ and $2$ to represent  the ground  states $X$, $Y$ and the first radially excited  states $X^\prime$, $Y^\prime$ respectively for simplicity.
 We rewrite the two   QCD sum rules in Eqs.\eqref{QCDST-1S}-\eqref{QCDST-2S} as
\begin{eqnarray}\label{QCDSR-I}
\lambda_1^2\exp\left(-\tau M_1^2 \right)&=&\Pi_{QCD}(\tau) \, ,
\end{eqnarray}
\begin{eqnarray}\label{QCDSR-II-re}
\lambda_1^2\exp\left(-\tau M_1^2 \right)+\lambda_2^2\exp\left(-\tau M_2^2 \right)&=&\Pi^{\prime}_{QCD}(\tau) \, ,
\end{eqnarray}
here we introduce the subscript $QCD$ to represent the QCD representation of the correlation functions $\Pi_{S/A/V/T}(p^2)$ below the continuum thresholds. We derive the QCD sum rules in Eq.\eqref{QCDSR-I} with respect to $\tau$ to obtain
the masses of the ground states,
\begin{eqnarray}\label{QCDSR-I-Dr}
M_1^2&=&\frac{D\Pi_{QCD}(\tau)}{\Pi_{QCD}(\tau)}\, .
\end{eqnarray}
We obtain the masses and pole residues of the ground states of the scalar, axialvector, vector and tensor $cc\bar{c}\bar{c}$ tetraquark states with the two coupled QCD sum rules shown in Eq.\eqref{QCDSR-I} and Eq.\eqref{QCDSR-I-Dr} \cite{WZG-QQQQ}.

Now we study the masses and pole residues of the first radial excited states. Firstly, let us derive  the QCD sum rules in Eq.\eqref{QCDSR-II-re} with respect to $\tau$ to obtain
\begin{eqnarray}\label{QCDSR-II-Dr}
\lambda_1^2M_1^2\exp\left(-\tau M_1^2 \right)+\lambda_2^2M_2^2\exp\left(-\tau M_2^2 \right)&=&D\Pi^{\prime}_{QCD}(\tau) \, .
\end{eqnarray}
From Eq.\eqref{QCDSR-II-re} and Eq.\eqref{QCDSR-II-Dr}, we can obtain the QCD sum rules,
\begin{eqnarray}\label{QCDSR-II-Residue}
\lambda_i^2\exp\left(-\tau M_i^2 \right)&=&\frac{\left(D-M_j^2\right)\Pi^{\prime}_{QCD}(\tau)}{M_i^2-M_j^2} \, ,
\end{eqnarray}
where the indexes $i \neq j$.
Then let us derive   the QCD sum rules in Eq.\eqref{QCDSR-II-Residue} with respect to $\tau$ to obtain
\begin{eqnarray}
M_i^2&=&\frac{\left(D^2-M_j^2D\right)\Pi_{QCD}^{\prime}(\tau)}{\left(D-M_j^2\right)\Pi_{QCD}^{\prime}(\tau)} \, , \nonumber\\
M_i^4&=&\frac{\left(D^3-M_j^2D^2\right)\Pi_{QCD}^{\prime}(\tau)}{\left(D-M_j^2\right)\Pi_{QCD}^{\prime}(\tau)}\, .
\end{eqnarray}
 The squared masses $M_i^2$ satisfy the  equation,
\begin{eqnarray}
M_i^4-b M_i^2+c&=&0\, ,
\end{eqnarray}
where
\begin{eqnarray}
b&=&\frac{D^3\otimes D^0-D^2\otimes D}{D^2\otimes D^0-D\otimes D}\, , \nonumber\\
c&=&\frac{D^3\otimes D-D^2\otimes D^2}{D^2\otimes D^0-D\otimes D}\, , \nonumber\\
D^j \otimes D^k&=&D^j\Pi^{\prime}_{QCD}(\tau) \,  D^k\Pi^{\prime}_{QCD}(\tau)\, ,
\end{eqnarray}
the indexes $i=1,2$ and $j,k=0,1,2,3$.
Finally we solve the equation analytically to obtain two solutions \cite{Baxi-G,WangZG-4430},
\begin{eqnarray}\label{QCDSR-II-M1}
M_1^2&=&\frac{b-\sqrt{b^2-4c} }{2} \, ,
\end{eqnarray}
\begin{eqnarray}\label{QCDSR-II-M2}
M_2^2&=&\frac{b+\sqrt{b^2-4c} }{2} \, .
\end{eqnarray}
From  the QCD sum rules in Eqs.\eqref{QCDSR-II-M1}-\eqref{QCDSR-II-M2}, we can obtain the masses of   both the ground states and the first radial excited states.
Both the QCD sum rules in Eq.\eqref{QCDSR-I-Dr} and Eq.\eqref{QCDSR-II-M1}  have one continuum threshold parameter,  both the continuum parameters $s_0$ and $s_0^\prime$ have uncertainties, in this aspect, the ground state masses from the QCD sum rules in Eq.\eqref{QCDSR-I-Dr} are not superior to the ones from Eq.\eqref{QCDSR-II-M1}. However, the   ground state masses from the QCD sum rules in Eq.\eqref{QCDSR-II-M1} suffer from additional uncertainties from the  first radial excited states.
In calculations, we observe that ground states masses from the QCD sum rules in Eq.\eqref{QCDSR-II-M1} underestimate  the experimental values \cite{Baxi-G,WangZG-4430}, so we neglect the QCD sum rules in Eq.\eqref{QCDSR-II-M1}.

\section{Numerical results and discussions}

We take the standard value of the gluon condensate 
\cite{SVZ79,Reinders85,ColangeloReview}, and  take the $\overline{MS}$ mass $m_{c}(m_c)=(1.275\pm0.025)\,\rm{GeV}$
 from the Particle Data Group \cite{PDG}.
We take into account
the energy-scale dependence of  the  $\overline{MS}$ mass from the renormalization group equation,
 \begin{eqnarray}
m_c(\mu)&=&m_c(m_c)\left[\frac{\alpha_{s}(\mu)}{\alpha_{s}(m_c)}\right]^{\frac{12}{25}} \, ,\nonumber\\
\alpha_s(\mu)&=&\frac{1}{b_0t}\left[1-\frac{b_1}{b_0^2}\frac{\log t}{t} +\frac{b_1^2(\log^2{t}-\log{t}-1)+b_0b_2}{b_0^4t^2}\right]\, ,
\end{eqnarray}
  where $t=\log \frac{\mu^2}{\Lambda^2}$, $b_0=\frac{33-2n_f}{12\pi}$, $b_1=\frac{153-19n_f}{24\pi^2}$, $b_2=\frac{2857-\frac{5033}{9}n_f+\frac{325}{27}n_f^2}{128\pi^3}$,  $\Lambda=213\,\rm{MeV}$, $296\,\rm{MeV}$  and  $339\,\rm{MeV}$ for the flavors  $n_f=5$, $4$ and $3$, respectively  \cite{PDG}. In this article, we choose the flavor number $n_f=4$ as we study the four-charm-quark states.

We should choose suitable continuum threshold parameters $s^\prime_0$ to avoid contaminations from the second radial excited states and borrow some ideas from the conventional charmonium states.
The  masses of the ground state, the first radial excited state and the second excited state  are $m_{J/\psi}=3.0969\,\rm{GeV}$,   $m_{\psi^\prime}=3.686097\,\rm{GeV}$ and $m_{\psi^{\prime\prime}}=4.039\,\rm{GeV}$ respectively from the Particle Data Group \cite{PDG}, the energy gaps are $m_{\psi^\prime}-m_{J/\psi}=0.59\,\rm{GeV}$, $m_{\psi^{\prime\prime}}-m_{J/\psi}=0.94\,\rm{GeV}$, we can  choose the continuum threshold parameters  $\sqrt{s_0^\prime}\leq M_{X/Y}+0.95\,\rm{GeV}$ tentatively and vary the continuum threshold parameters, energy scales of the QCD spectral densities and Borel parameters to satisfy
the  three   criteria:\\
$\bf 1.$ The ground state plus the first  radial excited state makes dominant contribution at the hadron  side;\\
$\bf 2.$ The operator product expansion is convergent below the continuum thresholds;\\
$\bf 3.$  The Borel platforms appear  both for the tetraquark masses and pole residues.

 In Ref.\cite{WZG-QQQQ}, we obtain the ground state masses of the scalar, axialvector, vector and tensor diquark-antidiquark type full-heavy tetraquark states with the QCD sum rules. In the present work, we take the ground state masses as the benchmark and study the masses of the excited states.
 After  trial and error,    we  reach the acceptable continuum threshold parameters, energy scales of the QCD spectral densities  and  Borel windows, which are shown in Table 1. From the Table, we can see that the pole dominance at the hadron side is well satisfied. In the Borel windows, the dominant contributions come from the perturbative terms,  the operator product expansion is well convergent.

\begin{figure}
 \centering
 \includegraphics[totalheight=5cm,width=7cm]{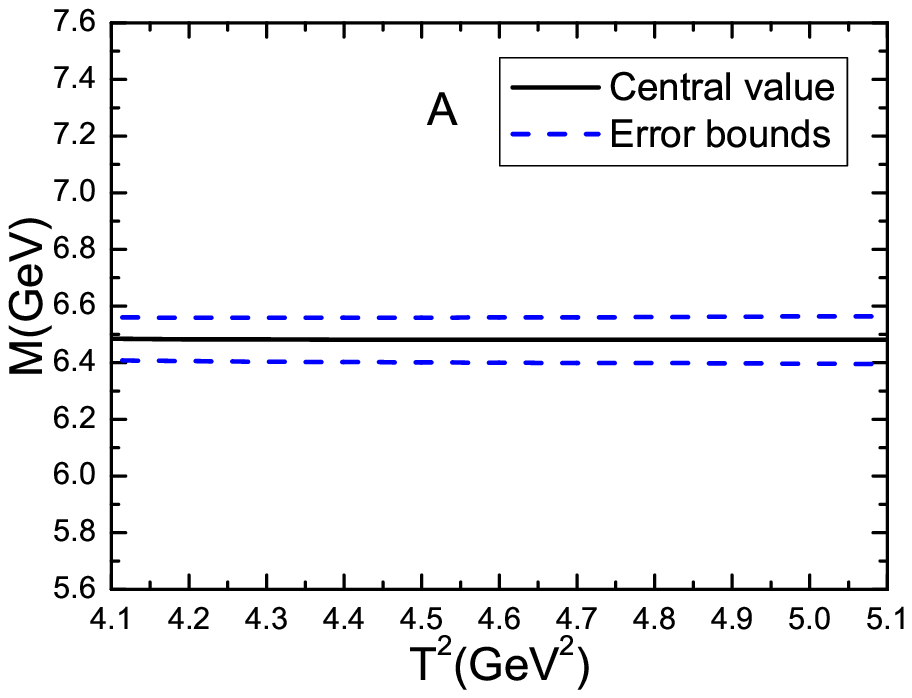}
 \includegraphics[totalheight=5cm,width=7cm]{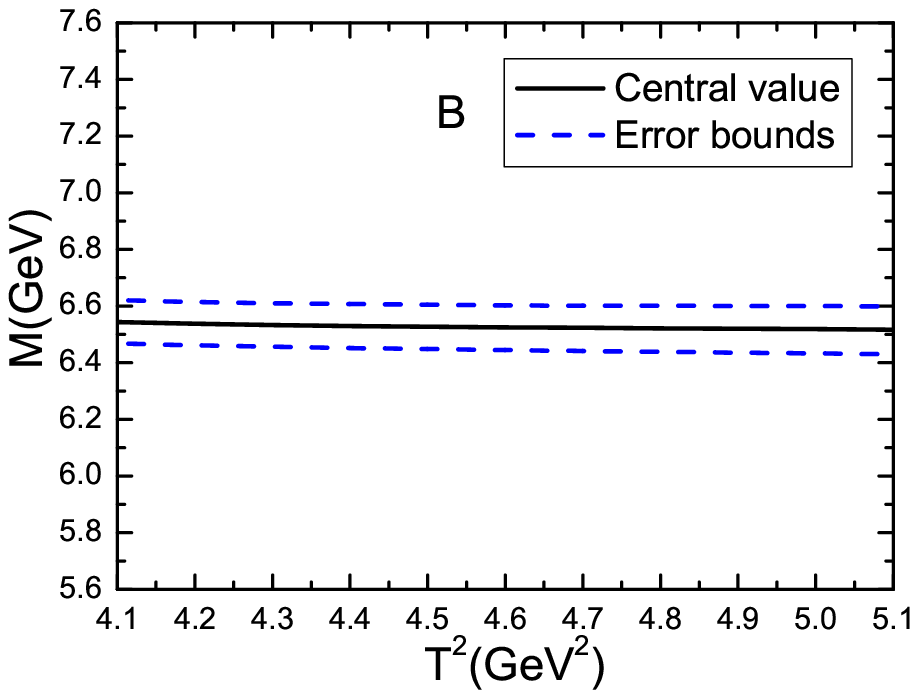}
 \includegraphics[totalheight=5cm,width=7cm]{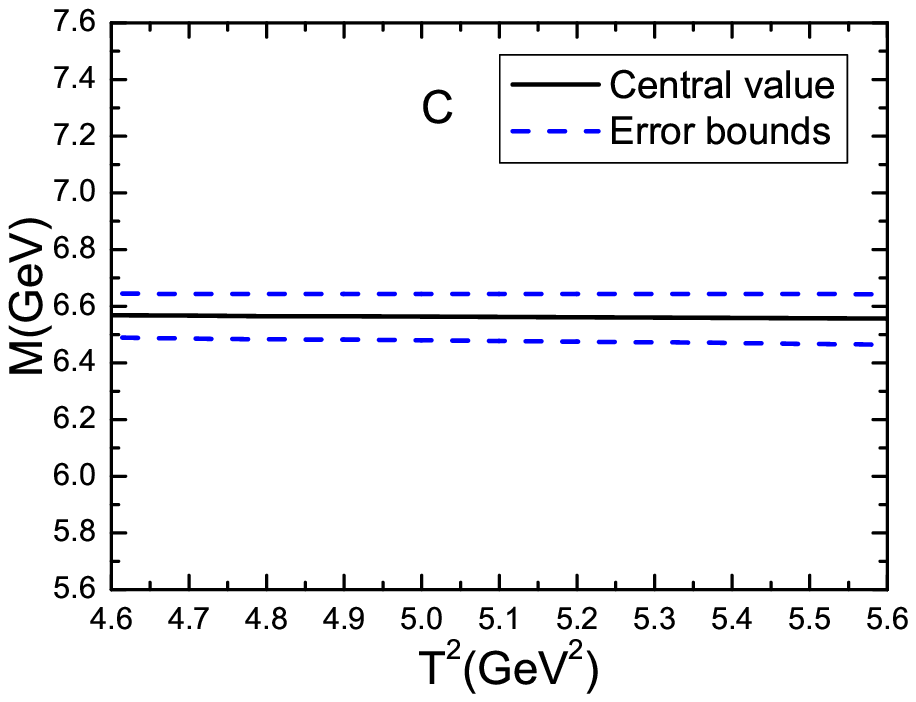}
 \includegraphics[totalheight=5cm,width=7cm]{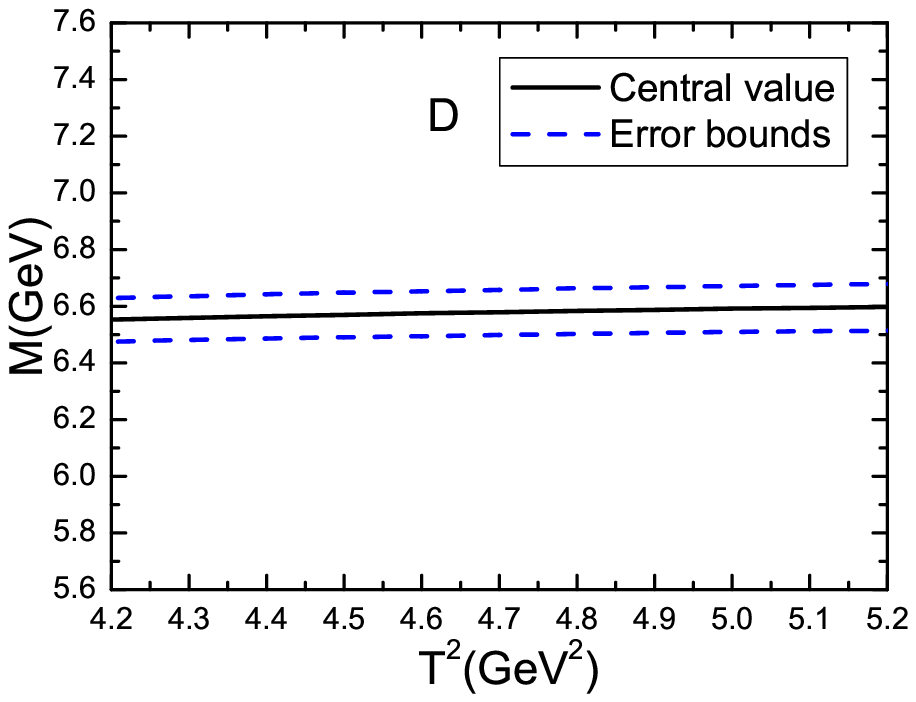}
         \caption{ The masses   of the first radial excited states of the tetraquark states  with variations of the Borel parameters $T^2$,  where the $A$, $B$, $C$ and $D$ denote the scalar, axialvector, vector and tensor tetraquark states, respectively.  }\label{mass-cccc}
\end{figure}

Now let us  take into account all uncertainties of the input parameters, and obtain the values of the masses and pole residues of the first radial excited states, which are also shown explicitly in Table 1 and Fig.\ref{mass-cccc}. The predicted masses and pole residues are rather stable with variations of the Borel parameters, the uncertainties originate from the Borel parameters in the Borel windows are very small, in other words, there appear Borel platforms. Now the three criteria are all satisfied, we expect to make reliable or sensible  predictions.

\begin{table}
\begin{center}
\begin{tabular}{|c|c|c|c|c|c|c|c|}\hline\hline
$J^{PC}$          &$T^2(\rm{GeV}^2)$  &$\sqrt{s_0}(\rm{GeV})$ &$\mu(\rm{GeV})$ &pole              &$M_{X/Y}(\rm{GeV})$ &$\lambda_{X/Y}(10^{-1}\rm{GeV}^5)$ \\ \hline

$0^{++}(\rm 2S)$  &$4.4-4.8$          &$6.80\pm0.10$          &$2.5$           &$(65-79)\%$       &$6.48\pm0.08$        &$7.41\pm1.12$  \\ \hline

$1^{+-}(\rm 2S)$  &$4.4-4.8$          &$6.85\pm0.10$          &$2.5$           &$(69-82)\%$       &$6.52\pm0.08$        &$5.56\pm0.80$  \\ \hline

$2^{++}(\rm 2S)$  &$4.9-5.3$          &$6.90\pm0.10$          &$2.5$           &$(63-76)\%$       &$6.56\pm0.08$        &$5.92\pm0.83$  \\ \hline

$1^{--}(\rm 2P)$  &$4.5-4.9$          &$6.90\pm0.10$          &$2.2$           &$(57-73)\%$       &$6.58\pm0.09$        &$3.46\pm0.58$  \\ \hline
\hline

\end{tabular}
\end{center}
\caption{ The Borel parameters, continuum threshold parameters, energy scales,  pole contributions, masses and pole residues of the $cc\bar{c}\bar{c}$  tetraquark states. }
\end{table}

\begin{table}
\begin{center}
\begin{tabular}{|c|c|c|c|c|c|c|c|}\hline\hline
$J^{PC}$                 &$M_{\rm 1}(\rm{GeV})$\cite{WZG-QQQQ}    &$M_{\rm 2}(\rm{GeV})$          &$M_{\rm 3}(\rm{GeV})$  \\ \hline

$0^{++} $                &$5.99\pm0.08$                           &$6.48\pm0.08$                  &$6.94\pm0.08$           \\ \hline

$1^{+-} $                &$6.05\pm0.08$                           &$6.52\pm0.08$                  &$6.96\pm0.08$           \\ \hline

$2^{++} $                &$6.09\pm0.08$                           &$6.56\pm0.08$                  &$7.00\pm0.08$          \\ \hline

$1^{--}$                 &$6.11\pm0.08$                           &$6.58\pm0.09$                  &$7.02\pm0.09$          \\ \hline
\hline

\end{tabular}
\end{center}
\caption{ The  masses of the $cc\bar{c}\bar{c}$ tetraquark states with the radial quantum numbers $n=1$, $2$ and $3$. }\label{mass-cccc-123}
\end{table}

 In Table \ref{mass-cccc-123}, we present the masses of the ground states and the first radial excited states from the QCD sum rules \cite{WZG-QQQQ}. If the masses of the ground states, the first radial excited states, the third radial excited states, etc satisfy the  Regge trajectory,
 \begin{eqnarray}
 M_n^2&=&\alpha (n-1)+\alpha_0\, ,
 \end{eqnarray}
 where the $\alpha$ and $\alpha_0$ are constants. We take the masses of the ground states and the first radial excited states shown in Table \ref{mass-cccc-123} as input parameters to fit the parameters $\alpha$ and $\alpha_0$, and obtain the masses of the second radial excited states, which are also shown in Table \ref{mass-cccc-123}.
 From the Table, we can see that the mass gaps $M_3-M_1=0.91\sim0.95\,\rm{GeV}$, which are consistent with the mass gap  $m_{\psi^{\prime\prime}}-m_{J/\psi}=0.94\,\rm{GeV}$. Furthermore, from the Table \ref{mass-cccc}, we can see that the continuum threshold parameters $\sqrt{s_0^\prime}\leq \overline{M}_3$, where the $\overline{M}_3$ represents  the central values of the masses of the second radial excited states.

From Table \ref{mass-cccc-123}, we can see that the predicted masses $M=6.48\pm0.08\,\rm{GeV}$, $6.52\pm0.08\,\rm{GeV}$, $6.56\pm0.08\,\rm{GeV}$
and $6.58\pm0.09\,\rm{GeV}$   for the first radial excited states of the scalar, axialvector, vector and tensor $cc\bar{c}\bar{c}$ tetraquark states are consistent with the broad structure above the threshold ranging from 6.2 to 6.8 GeV in the di-$J/\psi$ mass spectrum \cite{LHCb-cccc}, while the predicted masses $M=6.94\pm0.08\,\rm{GeV}$ and $6.96\pm0.08\,\rm{GeV}$ for the second radial excited states of the scalar and axialvector $cc\bar{c}\bar{c}$ tetraquark states are consistent with the narrow structure at about 6.9 GeV in the di-$J/\psi$ mass spectrum \cite{LHCb-cccc}. The present predictions  supports assigning the broad structure from 6.2 to 6.8 GeV in the di-$J/\psi$ mass spectrum to be the  first radial excited state of the scalar, axialvector, vector or tensor $cc\bar{c}\bar{c}$ tetraquark state, and assigning the narrow structure at about 6.9 GeV in the di-$J/\psi$ mass spectrum to be the  second radial excited state of the scalar or axialvector $cc\bar{c}\bar{c}$ tetraquark state.

\section{Conclusion}
In this article, we construct the scalar and tensor currents to  study the first radial excited states of the  scalar, axialvector, vector and tensor diquark-antidiquark type $cc\bar{c}\bar{c}$  tetraquark states with the QCD sum rules and obtain the masses and pole residues. Then we use the Regge trajectory to obtain the masses of the second radial excited states.  The present predictions  supports assigning the broad structure from 6.2 to 6.8 GeV in the di-$J/\psi$ mass spectrum to be the first radial excited state of the scalar, axialvector, vector or  tensor $cc\bar{c}\bar{c}$ tetraquark state, and assigning the narrow structure at about 6.9 GeV in the di-$J/\psi$ mass spectrum to be the  second radial excited state of the scalar or axialvector $cc\bar{c}\bar{c}$ tetraquark state.

\section*{Acknowledgements}
This  work is supported by National Natural Science Foundation, Grant Number  11775079.


\begin{thebibliography}{99}


\bibitem{Lloyd-2004} R. J. Lloyd and J. P. Vary, Phys. Rev. {\bf D70} (2004) 014009.

\bibitem{Barnea-2006} N. Barnea, J. Vijande and A. Valcarce, Phys. Rev. {\bf D73} (2006) 054004.

\bibitem{Berezhnoy-2012} A. V. Berezhnoy, A. V. Luchinsky and A. A. Novoselov, Phys. Rev. {\bf D86} (2012) 034004.

\bibitem{Heupel-2012}  W. Heupel, G. Eichmann and C. S. Fischer, Phys. Lett. {\bf B718} (2012) 545.



\bibitem{Bai-2016}  Y. Bai, S. Lu and J. Osborne, arXiv:1612.00012.

\bibitem{Richard-2017} J. M. Richard, A. Valcarce and J. Vijande, Phys. Rev. {\bf D95} (2017) 054019.

\bibitem{WZG-QQQQ} Z. G. Wang, Eur. Phys. J. {\bf C77} (2017) 432; Z. G. Wang and Z. Y. Di,  Acta Phys. Polon. {\bf B50} (2019) 1335.

\bibitem{Rosner-2017} M. Karliner, J. L. Rosner and S. Nussinov,  Phys. Rev. {\bf D95} (2017)  034011.

\bibitem{Chen-2017}  W. Chen, H. X. Chen, X. Liu, T. G. Steele and S. L. Zhu, Phys. Lett. {\bf B773} (2017) 247.

\bibitem{GuoFK-bbbb}  M. N. Anwar, J. Ferretti, F. K. Guo, E. Santopinto and B. S. Zou, Eur. Phys. J. {\bf C78} (2018)  647.


\bibitem{Polosa-2018} A. Esposito and A. D. Polosa, Eur. Phys. J. {\bf C78} (2018)  782.

\bibitem{Wu-2018} J. Wu, Y. R. Liu, K. Chen, X. Liu and S. L. Zhu, Phys. Rev. {\bf D97} (2018) 094015.


\bibitem{Hughes-2018}  C. Hughes, E. Eichten and C. T. H. Davies, Phys. Rev. {\bf D97} (2018)  054505.


\bibitem{Navarra-2019}    V. R. Debastiani and F. S. Navarra, Chin. Phys. {\bf C43} (2019)  013105.


\bibitem{Zhong-2019} M. S. Liu, Q. F. Lu, X. H. Zhong and Q. Zhao, Phys. Rev. {\bf D100} (2019)  016006.

\bibitem{Chen-2019} X. Chen,    arXiv:2001.06755.

\bibitem{Bedolla-2019} M. A. Bedolla, J. Ferretti, C. D. Roberts and E. Santopinto, arXiv:1911.00960.

 \bibitem{Ping-2020} C. Deng, H. Chen and J. Ping, arXiv:2003.05154.


\bibitem{Ohlsson-2020} P. Lundhammar and T. Ohlsson, arXiv:2006.09393.

\bibitem{Zhong-2020} M. S. liu, F. X. Liu, X. H. Zhong and Q. Zhao, arXiv:2006.11952.


\bibitem{LHCb-cccc} Liupan An [On behalf of the LHCb collaboration], Latest results on exotic hadrons at LHCb,  https://indico.cern.ch/event/900972/.


\bibitem{SVZ79}  M. A. Shifman, A. I. Vainshtein and V. I. Zakharov, Nucl. Phys. {\bf B147} (1979) 385;
Nucl. Phys. {\bf B147} (1979) 448.

\bibitem{Reinders85} L. J. Reinders, H. Rubinstein and S. Yazaki, Phys. Rept. {\bf 127} (1985) 1.



\bibitem{ColangeloReview} P. Colangelo and A. Khodjamirian, hep-ph/0010175.

\bibitem{PDG} M. Tanabashi et al, Phys. Rev. {\bf D98}   (2018) 030001.



\bibitem{Baxi-G} M. S. Maior de Sousa and R. Rodrigues da Silva, Braz. J. Phys. {\bf 46} (2016) 730.

\bibitem{WangZG-4430} Z. G. Wang, Commun. Theor. Phys. {\bf 63} (2015)  325;
Z. G. Wang, Chin. Phys. {\bf C44} (2020)  063105.

\end{thebibliography}
\end{document}